# Room-Temperature Superconductivity in Boron-Nitrogen Doped Lanthanum Superhydride

Yanfeng Ge,[1] Fan Zhang,[2,*] and Russell J. Hemley[3,4,*]

[1]*State Key Laboratory of Metastable Materials Science and Technology & Key Laboratory for Microstructural Material Physics of Hebei Province, School of Science, Yanshan University, Qinhuangdao, 066004, China*
[2]*Department of Physics, University of Texas at Dallas, Richardson, Texas 75080, USA*
[3]*Department of Physics, University of Illinois at Chicago, Chicago, Illinois 60607, USA*
[4]*Department of Chemistry, University of Illinois at Chicago, Chicago, Illinois 60607, USA*

*Corresponding authors: zhang@utdallas.edu; rhemley@uic.edu

Recent theoretical and experimental studies of hydrogen-rich materials at megabar pressures (*i.e.,* >100 GPa) have led to the discovery of very high-temperature superconductivity in these materials. Lanthanum superhydride $LaH_{10}$ has been of particular focus as the first material to exhibit a superconducting critical temperature ($T_c$) near room temperature. Experiments indicate that the use of ammonia borane as the hydrogen source can increase the conductivity onset temperatures of lanthanum superhydride to as high as 290 K. Here we examine the doping effects of B and N atoms on the superconductivity of $LaH_{10}$ in its fcc ($Fm\bar{3}m$) clathrate structure at megabar pressures. Doping at H atomic positions strengthens the $H_{32}$ cages of the structure to give higher phonon frequencies that enhance the Debye frequency and thus the calculated $T_c$. The predicted $T_c$ can reach 288 K in $LaH_{9.985}N_{0.015}$ within the average high-symmetry structure at 240 GPa.

**Introduction**

Realizing room-temperature superconductivity in hydrogen-rich materials under pressure is a major topic of great current interest. Specifically, high-pressure experiments motivated by density functional theory and Bardeen-Cooper-Schrieffer (BCS) theory have uncovered new classes of hydrogen-rich metal hydrides, or superhydrides, with superconducting critical temperatures ($T_c$'s) in the vicinity of room temperature at megabar pressures (*i.e.,* >100 GPa).[1,2] Calculations for the rare-earth hydrides predicted that $LaH_{10}$ and $YH_{10}$ would form dense hydride clathrate structures exhibiting $T_c$'s of 257-326 K at pressures of 200-300 GPa.[3,4]

X-ray diffraction experiments on the La-H system confirmed the formation and stability of the $LaH_{10}$ structure near the predicted pressures,[5] and subsequent electrical conductivity and critical current measurements confirmed the very high-temperature superconductivity of the phase.[6,7] Experiments that used ammonia borane ($NH_3BH_3$) as the hydrogen source indicated $T_c$'s beginning at 260 K, including conductivity onsets as high as 290 K that have been confirmed in a very recent work.[8] It was proposed that the high and variable $T_c$ arises from incorporation of N and/or B in the structure from the ammonia borane starting material.[7,8] Moreover, subsequent studies[9] that confirmed the reported structure[10] and high-temperature superconductivity[7,8] of $LaH_{10}$ found a slightly lower maximum $T_c$ of 250 K in experiments conducted without ammonia borane.

Using a method developed previously applied to $H_3S$,[11,12] here we confirm theoretically that B and N doping increases the $T_c$ of the lanthanum-based superhydride to room temperature. It is found that the doping at H atomic position makes the clathrate hydride structure more stable and hardens the low-frequency optical phonons. Although the electron-phonon coupling constant $\lambda$ is reduced, a marked increase of the logarithmically averaged phonon frequency $<\omega>_{log}$, equivalent to the rise of Debye temperature, enhances $T_c$. Remarkably, low-level doping of N in the H sublattice together with further increase in pressure raises the $T_c$ of lanthanum superhydride by at least 25 K, with a predicted $T_c$ for $LaH_{9.985}N_{0.015}$ at 240 GPa of 288 K.

**Results and Discussion**

Lanthanum superhydride, $LaH_{10}$, has been found to be the highest temperature superconducting phase so far in the La-H binary system. As a result, since the original theoretical prediction and experimental discovery,[3-7] the material has been the subject of numerous investigations (*e.g.*, Refs. [13-20]). Though the detailed properties of $LaH_{10}$, including anharmonicity and nuclear quantum effects, continue to be studied theoretically,[10,13-19] it is sufficient for the present purposes to examine its superconducting properties, and specifically, the nature of the electron-phonon coupling in the system, within a conventional quasiharmonic treatment of the lattice dynamics. The computational approach applied here is consistent with previous theoretical studies,[3,4] which have been largely confirmed by experimental results for lanthanum superhydride,[3-7] thereby validating the extensions of the techniques we use to examine the effects of doping, as described previously.[11,12]

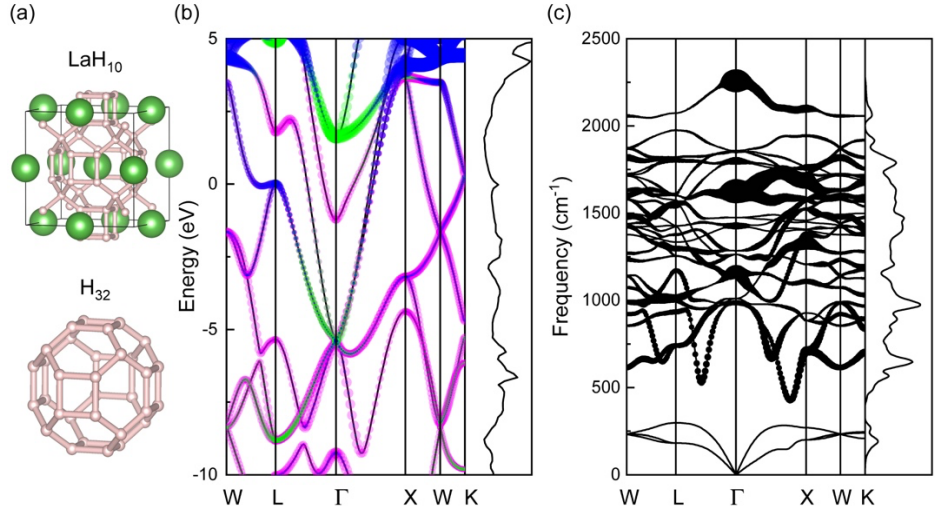

**Fig. 1.** (a) Optimized structure of LaH$_{10}$ with its fcc La sublattice and H$_{32}$ clathrate structure. (b) Electronic band structure and DOS of LaH$_{10}$ at 210 GPa. (c) Phonon dispersion and Eliashberg function $\alpha^2F(\omega)$ of LaH$_{10}$ at 210 GPa. The pink, blue, and green circles in (b) mark the projections of H-*s*, La-*d*, and La-*f* orbitals, respectively. The size of the black circles in (c) is proportional to the electron-phonon coupling linewidths. The results are consistent with the original predictions.[3]

The high-symmetry clathrate-type structure of LaH$_{10}$ consists of La atoms arrayed on an fcc lattice (space group $Fm\bar{3}m$), with a 32-atom hydrogen (H$_{32}$) cage surrounding each La atom containing six square and twelve hexagonal faces that are cross-linked by 8-hydrogen cubes [Fig 1(a)]. The central La atom acts as an electron source, donating electrons to the H sublattice, a result consistent with the ambient pressure ionization potential of La.[18] The electron transfer destabilizes molecular H$_2$ units in the structure that form in pure solid hydrogen at these pressures. This strong anion-cation interaction also enhances the stability of the H$_{32}$ cage, which is predicted to be dynamical unstable in La$_0$H$_{10}$.[19]

We focus on the properties of the material at pressures where the $T_c$ has a maximum for pure LaH$_{10}$, which corresponds to 210 GPa and close to the calculated quasiharmonic dynamical instability of the fcc structure.[3,10] The calculated nearest-neighbor H-H distances are 1.09 Å and 1.17 Å at this pressure, consistent with previous work.[3,4] Figure 1(b) shows that LaH$_{10}$ is a good metal, and the hole-like and electron-like bands around the L point generate a van Hove singularity near the Fermi energy.[18] This peak in the electronic density of states (DOS) near the Fermi level reaches 10.77 Hartree$^{-1}$/spin, and arises mainly from La-*d*, La-*f*, and H-*s* electrons. Phonon calculations reveal no imaginary frequencies indicating quasiharmonic dynamical stability of the $Fm\bar{3}m$ structure at 210 GPa [Fig. 1(c)]. On the other hand, at this pressure the structure is close to a dynamical instability, as indicated by the mode softening along Γ-L and Γ-X, which is associated with the maximum $T_c$ calculated at this level of theory for pure LaH$_{10}$.[3,10] The H-H distances lead to mixing of stretching and bending vibrations,[3] and phonon modes dominated by this H motion contribute to the electron-phonon coupling. As a result, the electron-phonon coupling, as reflected in the phonon linewidths, are distributed across many optical phonons [Fig. 1(c)]. The Allen-Dynes-modified McMillan formula predicted $T_c$ of $LaH_{10}$ at 210 GPa is 261 K with $\lambda$=2.96, $<\omega>_{log}$ =1090 K, and $\mu^*$=0.11. The calculated values agree with previous theoretical results and the $T_c$ is close to the experimental findings.[3,4,6,7] To examine the effect of doping we adopt the above value of $T_c$ = 261 K for pure $LaH_{10}$ at 210 GPa as a reference and use the same formalism for all the calculations while also examining the dependence of the results on $\mu^*$.

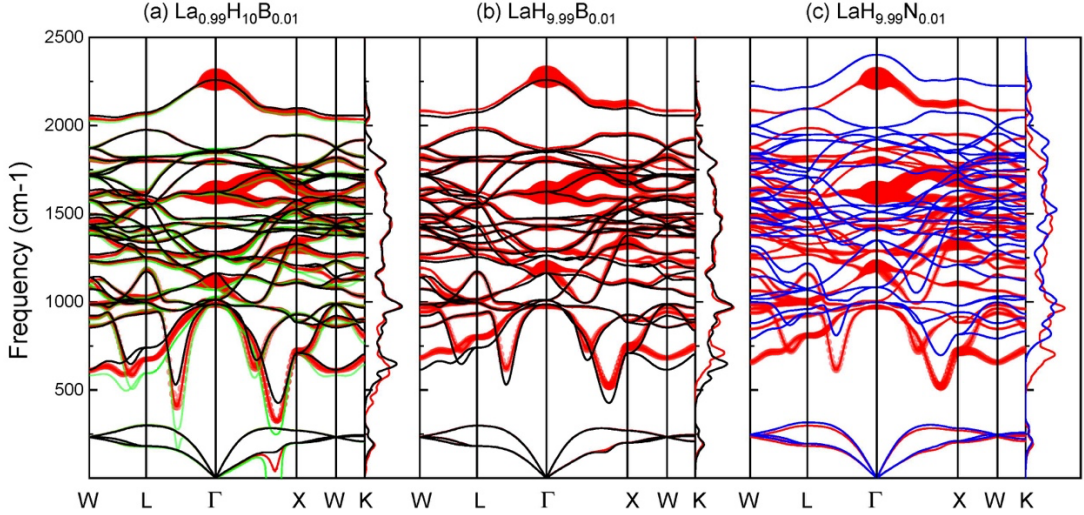

**Fig. 2.** (a) Phonon dispersion (red lines) and Eliashberg function $\alpha^2F(\omega)$ (red lines) of $La_{0.99}H_{10}B_{0.01}$, (b) $LaH_{9.99}B_{0.01}$ and (c) $LaH_{9.99}N_{0.01}$ at 210 GPa. The size of red circles is proportional to the electron-phonon coupling linewidths. In order to facilitate the comparison, the black lines in (a-b) show the phonon disperson and $\alpha^2F(\omega)$ of $LaH_{10}$ at 210 GPa. The green lines in (a) show the phonon dispersion of $La_{0.98}H_{10}B_{0.02}$ at 210 GPa with the imaginary modes along $\Gamma$-X. The blue lines in (c) show the phonon dispersion and $\alpha^2F(\omega)$ of $LaH_{9.99}N_{0.01}$ at 240 GPa.

Experimental studies have suggested that incorporation of N and/or B can improve the superconductivity of $LaH_{10}$.[8] Over the range of conditions explored to date, there is no evidence for major structural changes from cubic to higher pressure *P-T* structures. Thus, we consider the effects of modest doping of B and N, without structural change, on the electron distribution, structural stability, and superconductivity of the fcc $LaH_{10}$. We compare the results between $La_{0.99}H_{10}B_{0.01}$ and $LaH_{9.99}B_{0.01}$ at 210 GPa. Considering first the former, since B has many fewer valence electrons than La, the predominant effect of doping to create a virtual $La_{0.99}B_{0.01}$ atom is reduced electron transfer to the H atom framework. This weakens the $H_{32}$ cages as indicated by phonon softening shown in Fig. 2(a). Notably, frequencies of acoustic phonons along $\Gamma$-X become imaginary with modest doping, as indicated for $La_{0.98}H_{10}B_{0.02}$. By contrast, due to the increase in valence electrons of the virtual $H_{9.99}B_{0.01}$ atom, phonon hardening and notable changes in phonon frequencies around 500 cm$^{-1}$ occur [Fig. 2(b)]. The two types of doping result in phonon linewidths that are close to those of the LaH$_{10}$ prototype. This minor change in electron-phonon coupling arises from the similarity in the DOS at the Fermi level (10.57 and 10.80 Hartree$^{-1}$/spin for La$_{0.99}$H$_{10}$B$_{0.01}$ and LaH$_{9.99}$B$_{0.01}$, respectively), which is distinctly different from the doping effect found for H$_3$S.[11,12] Consequently, the peaks in the Eliashberg functions α$^2$F(ω) change little, though the peaks around 500 cm$^{-1}$ shift [Fig. 2(a) and (b)]. With the low-frequency optical phonon softening, λ of La$_{0.99}$H$_{10}$B$_{0.01}$ is enhanced but <ω>$_{log}$ is reduced. The calculated $T_c$ is 245 K with λ=3.33, <ω>$_{log}$=900 K, and μ*=0.11. By contrast, the effect is the opposite for LaH$_{9.99}$B$_{0.01}$, with λ=2.66 and <ω>$_{log}$ =1180 K, and the calculated $T_c$ is 269 K, *i.e.,* the $T_c$ is increased relative to LaH$_{10}$ and La$_{0.99}$H$_{10}$B$_{0.01}$ because of phonon hardening.

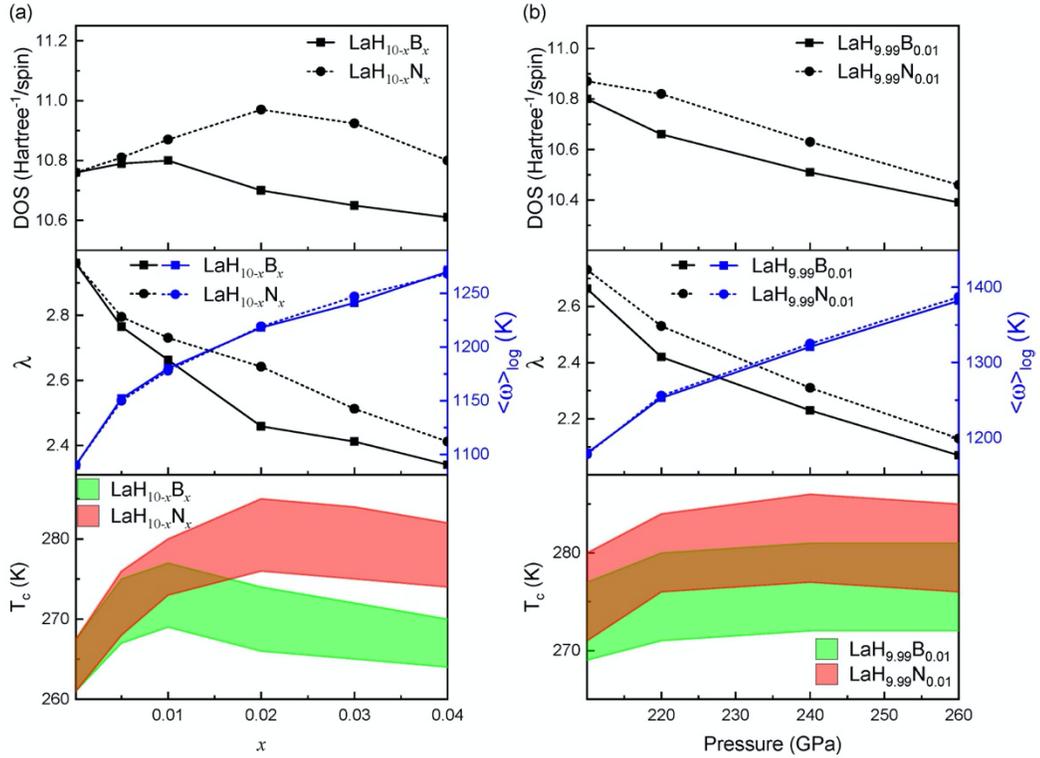

**Fig. 3.** (a) DOS, λ, <ω>$_{log}$, and $T_c$ of LaH$_{10-x}$B$_x$ and LaH$_{10-x}$N$_x$ with different doping levels at 210 GPa. (b) DOS, λ, <ω>$_{log}$, and $T_c$ of LaH$_{9.99}$B$_{0.01}$ and LaH$_{9.99}$N$_{0.01}$ at different pressures. For each $T_c$ curve, the upper and lower bounds are obtained by choosing μ*=0.10 and 0.11, respectively.

The above results indicate that superconductivity of fcc lanthanum superhydride is enhanced by increasing the stability of H$_{32}$ cage structure with additional charge transfer from the metal atom to the hydrogen framework. For this reason, we focus on the doping of B and N in the H sublattice (Fig. 3). Such doping results in small changes in the calculated DOS with doping level $x \leq 0.04$ for both LaH$_{10-x}$B$_x$ and LaH$_{10-x}$N$_x$, as shown in Fig. 3(a). Rising phonon frequencies after doping of B or N cause λ (<ω>$_{log}$) to decrease (increase), whereas the phonon linewidths show less of a change. According to the BCS theory, strong electron-phonon coupling and high Debye frequency increase $T_c$. It turns out that the increase of <ω>$_{log}$ in LaH$_{10-x}$B$_x$ and LaH$_{10-x}$N$_x$ can offset the negative effect of the lower λ and thus increase $T_c$. The larger DOS at the Fermi level gives LaH$_{10-}$ $_x$N$_x$ slightly stronger electron-phonon coupling and higher $T_c$ compared to LaH$_{10-x}$B$_x$ at a fixed doping level. Notably, the calculated $T_c$ of LaH$_{9.98}$N$_{0.02}$ is 276 (285) K with $\mu^*$=0.11 (0.10).[21]

We also studied the effect of pressure on the B and N doping in the hydrogen framework. Pressure compresses the H-H bonds, thereby increasing the frequencies of H atomic vibrations. As a result, there is an overall blue shift in the higher frequency portion of the Eliashberg function [Fig. 2(c)], which significantly differs from the effects of doping that largely changes the lower frequency optical phonons. Because of the hole-like and electron-like bands around the Fermi level, a weakly pressure-dependent DOS also gives rise to a small variation of phonon linewidths. As illustrated for LaH$_{9.99}$B$_{0.01}$ and LaH$_{9.99}$N$_{0.01}$ in Fig. 3(b), the effect of pressure is similar to that of doping and leads to a limited boost in $T_c$.

The dependence of $T_c$ on both doping level and pressure for LaH$_{10-x}$N$_x$ is shown in Fig. 4. The results indicate the predicted upper limit of $T_c$ by N doping while preserving the dynamically stable fcc LaH$_{10}$ structure. With $\mu^*$=0.10 (0.11), the $T_c$ is calculated to reach 288 (278) K for LaH$_{9.985}$N$_{0.015}$ at 240 GPa and $\lambda$=2.26 and $<\omega>_{\log}$ =1330 K. The results indicate an increase in $T_c$ of over 25 K compared to the maximum $T_c$ calculated at the same level of theory for LaH$_{10}$ (*i.e.*, 261 K at 210 GPa). Therefore, remarkably small doping levels are able to raise the calculated $T_c$ in the vicinity of room temperature.

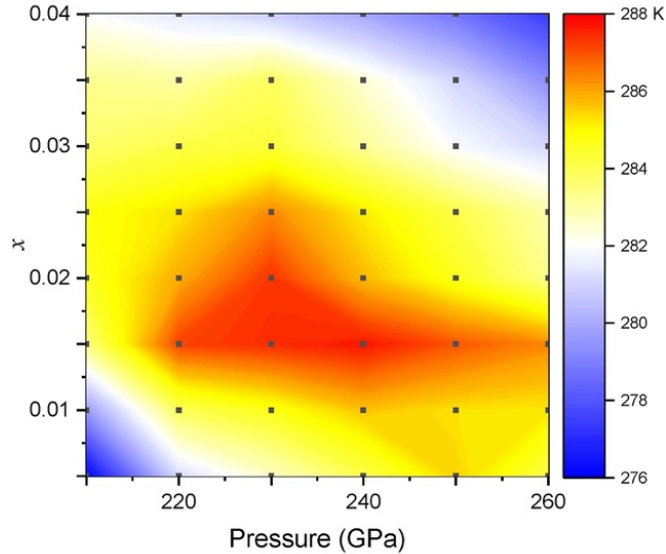

**Fig. 4.** Dependence of $T_c$ on doping level and pressure for LaH$_{10-x}$N$_x$ with $\mu^*$=0.10. The black dots represent the calculated data grid.

In summary, we have systematically studied the doping effects of B and N on the superconductivity of LaH$_{10}$ at megabar pressures, including the doping at La and H atomic positions. It is found that doping at the H atomic positions stabilizes the clathrate hydride structure and hardens the low-frequency optical phonons. Although the electron-phonon coupling constant λ is reduced, the marked increase in <ω>$_{log}$ (or effective Debye temperature) offsets the effect to increase the calculated $T_c$'s. As noted above, the effect of doping on $T_c$'s thus differs from that found previously for doping of $H_3S$ with P, Si, and C, which arose principally from changes in DOS at the Fermi level.[11,12] The results provide an explanation for the room-temperature superconductivity up to 290 K reported in several experiments on lanthanum superhydrides conducted using $NH_3BH_3$ as a hydrogen source.[6-8,20] A recently reported theoretical study of doping of $LaH_{10}$ suggested enhancing $T_c$ through DOS effects[22], similar to our predictions for $H_3S$.[11,12] By contrast, our calculations show that doping of $LaH_{10}$ enhances $T_c$ mainly by increasing the Debye frequency. Still higher $T_c$'s may be possible in the La-based system as a result of doping combined with structural changes induced by further application of pressure and temperature.[8] A related study of hole-doping by carbon in $H_3S$ causes a marked increase in $T_c$ to room temperature,[12] consistent with the maximum $T_c$ reported in a C-S-H mixture at 270 GPa.[23] Electron- or hole-doping with other elements in rare-earth and other hydrides is likely to lead to still higher $T_c$'s in hydrides as suggested by recent theoretical calculations on the Li-Mg-H system where stoichiometric Li was assumed[24] and the effects of proton dynamics on $T_c$ were not included.[25] Continued experimental and theoretical studies of doping effects and more complex chemical compositions of hydrides and related low-Z materials may lead ultimately very high-temperature superconductors that may be recoverable at or near ambient pressures.

**Methods**

The electron-phonon coupling was studied with the linear response theory and the Migdal-Eliashberg approach.[26-29] The Eliashberg electron-phonon spectral function reads

$$\alpha^2 F(\omega) = \frac{1}{2\pi N(\epsilon_F)} \sum_{q,\nu} \frac{\gamma_{q,\nu}}{\omega_{q,\nu}} \delta(\omega - \omega_{q,\nu}), \quad (1)$$

where $N(\epsilon_F)$ is the DOS per spin at the Fermi level and $\gamma_{q,\nu}$ is the phonon linewidth given by the Fermi's golden rule,

$$\gamma_{q,\nu} = 2\pi \omega_{q,\nu} \sum_{i,j,k} \left| M^{q,\nu}_{ik,jk+q} \right|^2 \delta(\epsilon_{ik} - \epsilon_F) \delta(\epsilon_{jk+q} - \epsilon_F). \quad (2)$$

The microscopic electron-phonon matrix element $M^{q,\nu}_{ik,jk+q}$ describes the scattering of an electron at the Fermi surface from a state of momentum **k** to another state of **k+q** perturbed by a phonon mode (**q**,ν). In the electron-phonon coupling model, the superconducting state is obtained by solving the Eliashberg gap equation with the electron-phonon and electron-electron interactions in the normal state. Describing the manifestation of electron-phonon interaction in superconductivity, the electron-phonon coupling constant λ was obtained from $\lambda = 2 \int \alpha^2 F(\omega) \omega^{-1} d\omega$.

The electron-electron Coulomb interaction was represented by an effective constant $\mu^*$. The $T_c$ was estimated by the Allen-Dynes-modified McMillan formula,[29]

$$T_c = f_1 f_2 \frac{<\omega>_{log}}{1.20} \exp\left[-\frac{1.04(1+\lambda)}{\lambda - \mu^*(1+0.62\lambda)}\right], \quad (3)$$

where $f_1$ and $f_2$ are the strong coupling and shape correction factors, respectively, and $\langle\omega\rangle_{\log}$ is the logarithmically averaged phonon frequency.

Technical details of computation are as follows. Density functional theory was implemented within the generalized gradient approximation using the Perdew-Burke-Ernzerh method in the ABINIT package.[30-32] The ion and electron interactions were treated with the Hartwigsen-Goedecker-Hutter pseudopotentials.[33] The basis set containing all plane waves up to the cutoff energy of 30 Hartree and the zone-centered Monkhorst-Pack $k$-mesh of $24\times24\times24$ were used in all calculations. The phonon spectra and electron-phonon coupling were calculated on an $8\times8\times8$ $q$-grid using the density functional perturbation theory.[34] For electron-phonon coupling, the matrix elements were determined by second-order perturbation theory. To improve accuracy we performed a Fourier interpolation of the matrix elements to obtain them on a finer $q$-grid identical to the $k$-mesh. The doping of B and N atoms was simulated by the self-consistent virtual crystal approximation, where the pseudopotentials of $A_{1-x}B_x$ are given by $(1-x)V_A + xV_B$.

**Acknowledgments**

This work was supported by National Natural Science Foundation of China (Grants No.11904312), the Project of Hebei Educational Department of China (Grants No. QN2018012), the UT Dallas Research Enhancement Fund, the US National Science Foundation (Grant No. DMR-1933622), and the US Department of Energy (Grant Nos. DE-SC0020340 and DE-NA0003975).

**References**

1. Bi, T., Zafiri, N., Terpstra, T., and Zurek, E. The search for superconductivity in high pressure hydrides. *Reference Chemistry Molecular Sciences and Chemical Engineering*, 1-36 (2019).
2. Flores-Livas, J. A., Boeri, L., Sanna, A., Profeta, G., Arita, R., and Eremets, M. A perspective on conventional high-temperature superconductors at high pressure: Methods and materials. *Phys. Rep.* **856**, 1-78 (2020).
3. Liu, H., Naumov, I. I., Hoffmann, R., Ashcroft, N. W., and Hemley, R. J. Potential high-Tc superconducting lanthanum and yttrium hydrides at high pressure. *Proc. Natl. Acad. Sci.* **114**, 6990-6995 (2017).
4. Peng, F., Sun, Y., Pickard, C. J., Needs, R. J., Wu, Q., and Ma, Y. Hydrogen clathrate structures in rare earth hydrides at high pressures: Possible route to room-temperature superconductivity. *Phys. Rev. Lett.* **119**, 107001 (2017).
5. Geballe, Z. M., Liu, H., Mishra, A. K., Ahart, M., Somayazulu, M., Meng, Y., Baldini, M., and Hemley, R. J. Synthesis and stability of lanthanum superhydrides. *Angew. Chem. Inter. Ed.* **57**, 688-692 (2018).
6. Somayazulu, M., Ahart, M., Mishra, A. K., Geballe, Z. M., Baldini, M., Meng, Y., Struzhkin, V. V., and Hemley, R. J. Evidence for superconductivity above 260 K in lanthanum superhydride at megabar pressures. *Phys. Rev. Lett.* **122**, 027001 (2019).
7. Hemley, R. J., Ahart, M., Liu, H., and Somayazulu, M. Road to room-temperature superconductivity: $T_c$ above 260 K in lanthanum superhydride under pressure in *Superconductivity and Pressure: A Fruitful Relationship on the Road to Room Temperature Superconductivity*, Madrid, Spain, May 21-22, 2018, edited by M. A. Alario-Franco. (Fundación Ramón Areces, Madrid), 199-213.

8. Grockowiak, A. D., Ahart, M., Helm, T., Coniglio, W. A., Kumar, R., Somayazulu, M., Meng, Y., Oliff, M., Williams, V., Ashcroft, N. W., Hemley, R. J., and Tozer, S. W. Possible hot hydride superconductivity above 550 K. arXiv:2006.03004.

9. Drozdov, A. P., Kong, P. P., Minkov, V. S., Besedin, S. P., Kuzovnikov, M. A., Mozaffari, S., L. Balicas, Balakirev, F., Graf, D., Prakapenka, V. B., Greenberg, E., Knyazev, D. A., Tkacz, M., and Eremets, M. I. Superconductivity at 250 K in lanthanum hydride under high pressures. *Nature* **569**, 528-531 (2019).

10. Liu, H., Naumov, I. I., Geballe, Z. M., Somayazulu, M., Tse, J. S., and Hemley, R. J. Dynamics and superconductivity in compressed lanthanum superhydride. *Phys. Rev. B.* **98**, 100102 (2018).

11. Ge, Y., Zhang, F., and Yao, Y. First-principles demonstration of superconductivity at 280 K in hydrogen sulfide with low phosphorus substitution. *Phys. Rev. B* **93**, 224513 (2016).

12. Ge, Y., F Zhang, R. P. Dias, R. J. Hemley, and Y. Yao, Hole-doped room-temperature superconductivity in $H_3S_{1-x}Z_x$ (Z=C, Si). *Mater. Today Phys.*, 15, 100330 (2020).

13. Kruglov, I. A., Semenok, D. V., Song, H., Szczesniak, R., Wrona, I. A., Akashi, R., Esfahani, M. M. D., Duan, D., Cui, T., Kvashnin, A. G., and Oganov, A. R. Superconductivity of $LaH_{10}$ and $LaH_{16}$ polyhydrides. *Phys. Rev. B* **101**, 024508 (2020).

14. Errea, I., Belli, F., Monacelli, L., Sanna, A., Koretsune, T., Tadano, T., Bianco, R., Calandra, M., Arita, R., Mauri, F., and Flores-Livas, J. A. Quantum crystal structure in the 250-kelvin superconducting lanthanum hydride. *Nature* **578**, 66-72 (2019).

15. Shipley, A. M., Hutcheon, M. J., Johnson, M. S., Needs, R. J., and Pickard, C. J. Stability and superconductivity of lanthanum and yttrium decahydrides. *Phys. Rev. B* **101**, 224511 (2020).

16. Sun, W., Kuang, X., Keen, H. D. J., Lu, C., and Hermann, A. Second group of high-pressure high-temperature lanthanide polyhydride superconductors. *Phys. Rev. B* **102**, 144524 (2020).

17. Sun, D., Minkov, V. S., Mozaffari, S., Chariton, S., Prakapenka, V. B., Eremets, M. I., Balicas, l., and Balakirev, F. F. High-temperature superconductivity on the verge of a structural instability in lanthanum superhydride. arXiv:2010.00160.

18. Wang, C., Yi, S., and Cho, J.-H. Pressure dependence of the superconducting transition temperature of compressed $LaH_{10}$. *Phys. Rev. B* **100**, 060502 (2019).

19. Yi, S., Wang, C., Jeon. H., and Cho, J.-H. Stabilization mechanism of clathrate H cages in a room-temperature superconductor $LaH_{10}$. arXiv:2007.01531.

20. Hong, F., Yang, L., Shan, P., Yang, P., Liu, Z., Sun, J., Yin, Y., Yu, X., Cheng, J. and Zhao, Z. Superconductivity of lanthanum superhydride investigated using the standard four-probe configuration under high pressures. *Chin. Phys. Lett.* **37**, 107401 (2020).

21. *The Electron-Phonon Interaction in Metals*, Selected Topics in Solid State Physics Vol. 16, edited by E. P. Wohlfahrt (North-Holland, Amsterdam, 1981).

22. Flores-Livas, J. A., Wang, T., Nomoto, T., Koretsune, T., Ma, Y., Arita, R., and Eremets, M. Reaching room temperature superconductivity by optimizing doping in $LaH_{10}$. arXiv:2010.06446.

23. Snider, E., Dasenbrock-Gammon, N., McBride, R., Debessai, M., Vindana, H., Vencatasamy, K., Lawler, K. V., Salamat, A., and Dias, R. P. Room-temperature superconductivity in a carbonaceous sulfur hydride. *Nature* **586**, 373–377 (2020).

24. Sun, Y., Lv, J., Xie, Y., Liu, H., and Ma, Y. Route to a superconducting phase above room temperature in electron-doped hydride compounds under high pressure. *Phys. Rev. Lett.* **123**, 097001 (2019).
25. Wang, H., Yao, Y., Peng, F., Hemley, R. J., and Liu, H. Quantum and classical proton diffusion in superconducting clathrate hydrides. *Phys. Rev. Lett.*, submitted.
26. McMillan, W. L. Transition temperature of strong-coupled superconductors. *Phys. Rev.* **167**, 331 (1968).
27. Allen, P. B. Neutron spectroscopy of superconductors. *Phys. Rev. B* **6**, 2577 (1972).
28. Allen, P. B., and Silberglitt, R. Some effects of phonon dynamics on electron lifetime, mass renormalization, and superconducting transition temperature. *Phys. Rev. B* **9**, 4733 (1974).
29. Allen, P. B., and Dynes, R. C. Transition temperature of strong-coupled superconductors reanalyzed. *Phys. Rev. B* **12**, 905 (1975).
30. Gonze, X., Amadon, B., Anglade, P.-M., Beuken, J.-M., et al. ABINIT: First-principles approach to material and nanosystem properties. *Comput. Phys. Commun.* **180**, 2582 (2009).
31. Gonze, X., Jollet, F., Abreu Araujo, F., Adams, D., et al. Recent developments in the ABINIT software package. *Comput. Phys. Commun.* **205**, 106 (2016).
32. Gonze, X., Amadon, B., Antonius, G., Arnardi, F., et al. The Abinit project: Impact, environment and recent developments. *Comput. Phys. Commun.* **248**, 107042 (2020).
33. Hartwigsen, C., Goedecker, S., and Hutter, J. Relativistic separable dual-space Gaussian pseudopotentials from H to Rn. *Phys. Rev. B* **58**, 3641 (1998).
34. Baroni, S., Gironcoli, S. D., Corso, A. D., and Giannozzi, P. Phonons and related crystal properties from density-functional perturbation theory. *Rev. Mod. Phys.* **73**, 515 (2001).